# Battling Botpoop using GenAI for Higher Education: A Study of a Retrieval Augmented Generation Chatbot's Impact on Learning

Maung Thway[1*], Jose Recatala-Gomez[2*], Fun Siong Lim[3], Kedar Hippalgaonkar[2,4], Leonard W. T. Ng[2#]

[1]Centre for the Applications of Teaching & Learning Analytics for Students (ATLAS), Nanyang Technological University, Singapore 639798, Singapore.

[2]School of Materials Science and Engineering, Nanyang Technological University, Singapore 639798, Singapore.

[3]Centre for IT Services (CITS), Nanyang Technological University, Singapore 639798, Singapore.

[4]Institute of Materials Research and Engineering, Agency for Science, Technology and Research, 2 Fusionopolis Way, 138634, Singapore

* These authors contributed equally

# Author to whom the correspondence should be addressed

## Abstract

Generative artificial intelligence (GenAI) and large language models (LLMs) have simultaneously opened new avenues for enhancing human learning and increased the prevalence of poor-quality information in student response - termed 'Botpoop'. This study introduces Professor Leodar, a custom-built, Singlish-speaking Retrieval Augmented Generation (RAG) chatbot designed to enhance educational while reducing Botpoop. Deployed at Nanyang Technological University, Singapore, Professor Leodar offers a glimpse into the future of AI-assisted learning, offering personalized guidance, 24/7 availability, and contextually relevant information. Through a mixed-methods approach, we examine the impact of Professor Leodar on learning, engagement, and exam preparedness, with 97.1% of participants reporting positive experiences. These findings help define possible roles of AI in education and highlight the potential of custom GenAI chatbots. Our combination of chatbot development, in-class deployment and outcomes study offers a benchmark for GenAI educational tools and is a stepping stone for redefining the interplay between AI and human learning.

## Introduction

The rapid advancements in Generative Artificial Intelligence (GenAI) and its increasing integration into educational settings are profoundly reshaping the landscape of human learning.[1] As large language models (LLMs) like GPT-3.5 and GPT-4[2,3], demonstrate remarkable capabilities in knowledge retrieval and generation, educators and institutions grapple with how to effectively harness these tools to enhance student learning while navigating complex challenges around equity, academic integrity, and student skills development.[4,5] Already, these technologies have significantly impacted higher education, leading to varied responses from academic institutions, ranging from acknowledgment of its potential benefits.[6–8] to strict prohibitions and firewalls in some Australian schools.[9,10]

GenAI-powered chatbots, while highly interactive and adept at general knowledge retrieval, often struggle to provide reliable domain-specific expertise when applied in specialized fields such as healthcare[11], finance[12], and law.[13] This limitation underscores the critical need for carefully tailored GenAI solutions in educational contexts, where in-depth subject knowledge and pedagogical adaptations are essential for effectively supporting student learning. As stated by Hannigan *et al.* *"*This is of utmost importance if we are to prevent the proliferation of "bot****", which are those chatbots that generate untruthful content used by humans for tasks.[14]" Here, we adopt Hannigan *et al's* term in human education as 'Botpoop', which refers to inaccurate information produced by open-source chatbots.

While prior research has explored applications of AI tutoring systems[15,16], and chatbots in specific domains such as mathematics[17], foreign languages[18], and engineering[19–21] there remain significant gaps in understanding around designing and deploying GenAI chatbots that are carefully tailored to the learning needs and contexts of specific student populations. Colace *et al.* developed a chatbot to act as an "*e-Tutor*" or a "*study buddy*" during student e-learning activities[20], but their evaluation was limited to post-exam surveys of students who passed the courses.

In another study, Abedi *et al.*[21] used ChatGPT to develop a chatbot for a fluid mechanics course, finding that carefully designed prompts and integrations with external tools like Wolfram could yield highly accurate responses to student queries. However, these studies did not comprehensively examine the impact of the chatbots on students' actual learning experiences and outcomes. Dan *et al.*[22] developed EduChat, a personalized LLM-based chatbot fine-tuned on educational texts and expert-curated instructions. While effective in open Q&A and Socratic teaching, EduChat required substantial resources, highlighting the challenges of creating tailored GenAI solutions for education. A state university in Turkey deployed the use of ChatGPT in a small class with a limited, single-modal survey to gauge its effectiveness.[23] This study confirmed all educator biases with concerns that open-source GenAI might be considered 'the easy way out' for students and that a highly tailored solution that gave accurate answers is desired.[23,24] To address these gaps, we present a novel study on the development and deployment of a custom GenAI chatbot, Prof. Leodar, to enhance human learning for undergraduate engineering students at the Nanyang Technological University in Singapore.[24] Our study focuses on three key research questions (RQs):

- **RQ1:** How effective is Professor Leodar in enhancing students' learning outcomes, engagement, and exam preparedness?
- **RQ2:** How does Professor Leodar compare to other GenAI tools in supporting student learning?
- **RQ3:** What improvements can be made to optimize Professor Leodar's effectiveness as an educational tool?

Unlike typical approaches that fine-tune LLMs on static datasets[25] , Prof. Leodar leverages Retrieval Augmented Generation[26] to provide students with contextually relevant information grounded in their specific course content, updated in real-time. We employ a robust mixed-methods approach[27–31] combining analytics, surveys, and focus group discussions to comprehensively evaluate the impact of Prof. Leodar on students' learning experiences and outcomes.

Our results demonstrate that interacting with Prof. Leodar significantly enhanced students' perceived and actual learning, with the chatbot serving as a highly engaging 24/7 tutor that provided personalized guidance and facilitated self-directed exploration, which positively reduced the prevalence of Botpoop in final examinations and assignments. We identify key factors influencing the effectiveness of the chatbot, such as the quality and specificity of the retrieval-augmented outputs, the use of scaffolded questioning strategies, and the availability of multimodal interaction features. Notably, we find that the chatbot's use of casual language and a relatable persona played a crucial role in fostering engagement and positive perceptions of the learning support.

The insights from this study inform the design of GenAI technologies as powerful tools for augmenting and enhancing human learning, with broad applications across diverse educational settings. Our findings highlight the importance of domain-specific customization, real-time knowledge integration, and persona design in creating effective AI-powered learning companions. As educational institutions increasingly adopt GenAI, this work provides a foundation for evidence-based practices to optimally integrate these tools while proactively addressing key pedagogical and ethical considerations. Ultimately, this research significantly advances the frontiers of human-AI collaborative learning and charts new directions for educational technologies that adapt to and empower diverse learners.

## Methods

### Building Prof. Leodar

We developed Prof. Leodar, a custom GenAI chatbot, to provide personalized learning support for students in the "MS0003: Introduction to Data Science and Artificial Intelligence" course at Nanyang Technological University, Singapore. The chatbot's architecture (Figure 1) leverages Retrieval Augmented Generation (RAG)[26], a technique that combines pre-trained language models with dynamic information retrieval to generate contextually relevant responses.

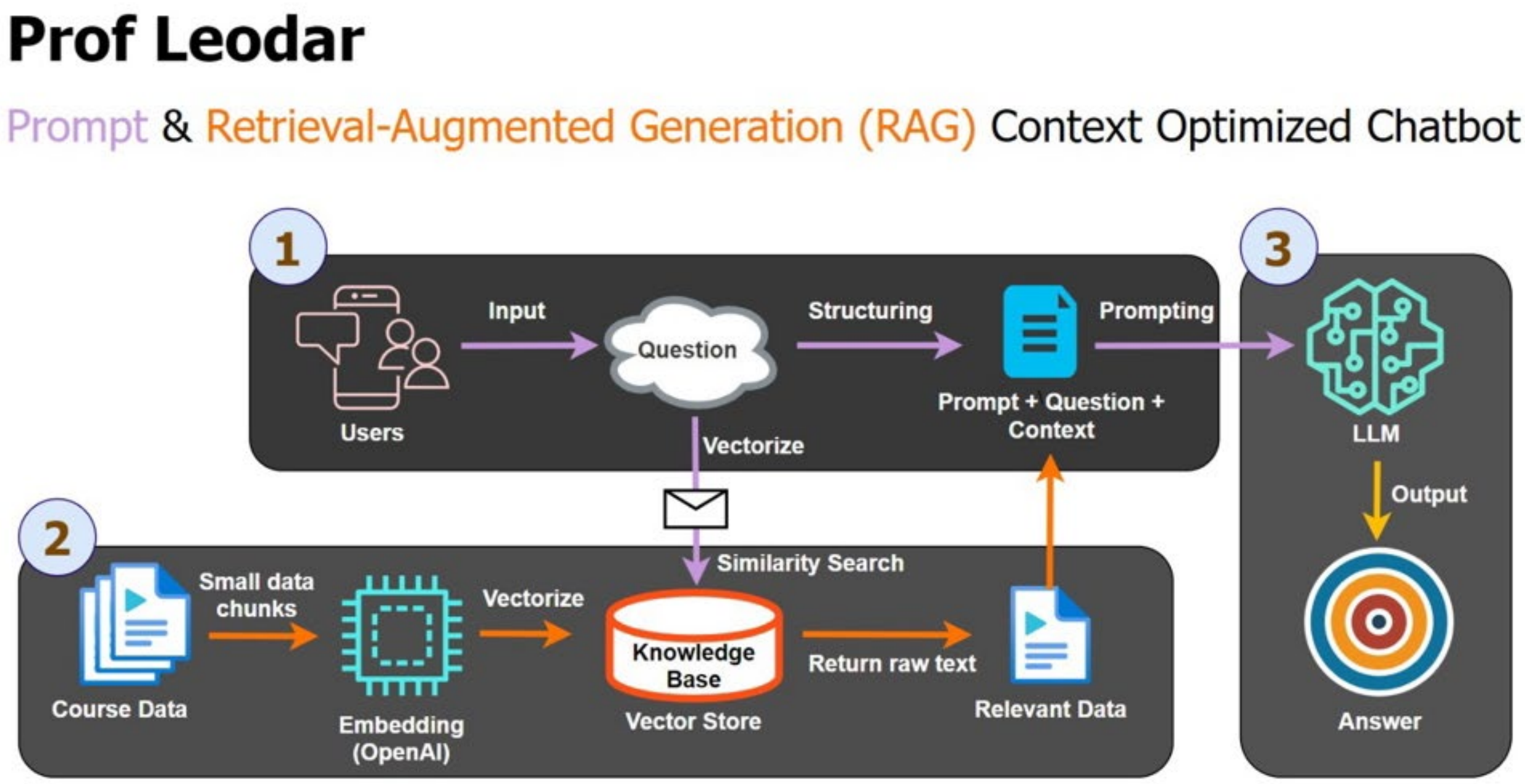

**Figure 1.** Overall architecture and workflow of the chatbot Prof. Leodar.

To create the chatbot's knowledge base, we curated a diverse corpus encompassing course materials (lecture notes, Jupyter notebooks), and domain-specific textbooks (Supplementary Information (SI) Section 1). Our approach employs Retrieval-Augmented Generation (RAG), designed to improve the quality of the response and relevance.

The knowledge base was segmented and embedded using OpenAI's embedding model, with semantically similar chunks stored in a Vector Store for efficient retrieval. When a student asks a question, Prof. Leodar transforms it into a numerical form (embedding) and searches for similar segments within the Vector Store. The most pertinent segments are then used as context for generating the response. Prof. Leodar then utilizes Anthropic's Claude 3 model to interpret these segments and generate human-like responses by seamlessly integrating this retrieved information. The combination of RAG and the Claude 3 model enables Prof. Leodar to provide accurate, relevant, and contextually appropriate responses to student inquiries.

We chose RAG over alternative methods, such as fine-tuning an open-source language model (LLM), for several reasons. First, RAG allows the model to access up-to-date and domain-specific information without the need for retraining. This is particularly important in our context, where course materials may be updated frequently. Second, RAG reduces the risk of the model generating inaccurate or fabricated answers (known as hallucination) by grounding the responses in verified information from the knowledge base. While RAG does not completely eliminate the issue, it significantly reduces the chance of generating deceptive answers compared to standard.

A series of metaprompts were defined to provide Prof. Leodar with a "persona". This prompts also serve as guidelines for Prof. Leodar to provide responses:

- Your name is Prof Leodar. You are a helpful, polite, fact-based assistant who cracks lame jokes in Singlish.
- You are a personalised helper for the course MS0003 Data Science and Artificial Intelligence. The module code is MS0003.

- Your job is to help them answer questions about the course and course content. You should be able to deal with course-related enquiries, both administrative and content-wise. You will answer the question in a brief and simple manner. Please provide example code when asked.
- If someone asks you what you are powered by, honestly tell them that you are on steroids and powered by Claude v3.
- If the student asks you about non-academic related stuff, make a sarcastic joke and then politely guide the student to only chat with you about academic related content.
- Firstly, you should refer to the history of conversation between you and the student. Then, you may use the content of the "system" provided before the "user" below if it is useful to answer the question.

A key feature of Prof. Leodar is its ability to adapt to students' evolving needs through real-time updates to its knowledge base. Weekly homework exercises and solutions were continuously added to the corpus, ensuring that the chatbot could provide timely and relevant support throughout the course. Complete code to replicate this work is given in SI-Production Code.

## Costs and Deployment

The costs associated with developing and deploying Prof. Leodar can be categorized into two main phases: the exploratory phase, during which various architectures were tested, and the deployment and maintenance phase. During the exploratory phase, a total of USD 523.04 was incurred, primarily for AWS services such as SageMaker (USD 85.30) and other resources like S3 Buckets and web hosting. Initially, Prof. Leodar utilized the GPT-4 as LLM, resulting in a cost of USD 504.36. However, to improve response times and reduce costs, we transitioned to Claude 3 on March 11th, 2024, resulting in a reduced LLM cost of USD 151.24 for the remainder of the course. The average daily cost of running Prof. Leodar for 154 users was USD 16.01, translating to a monthly cost of USD 3.11 per user (Table 1). Prof. Leodar was deployed from Week 2 of the MS0003 course in the academic year 2024/2025 onwards. During the entire deployment, a total of 12,334 questions were presented to Prof. Leodar by 154 users which is an average of 81 questions per user (SI Annex F).

**Table 1.** Benchmark of Prof. Leodar with other popular chatbots cloud services for personalised education.

| Chatbot | Personalization* | Transferability# | Operation Cost in USD (per 100 users) |
|---|---|---|---|
| ChatGPT | None | Low | 2,000 |
| Azure | High | High | 7,714 |
| Google Cloud Platform | High | High | 5,714 |
| Prof. Leodar (this work) | High | High | 182 |

*Personalization: ability to tailor chatbot framework to desired specifications.

#Transferability: ability of the chatbot framework to be adapted to a different application.

## Evaluating Human Learning

To comprehensively assess Prof. Leodar's impact on human learning, we perform a mixed-methods study[27,31,32] that is known to provide comprehensive and nuanced understanding of complex research questions by capitalizing on the respective strengths of quantitative (e.g. generalizability, deductive reasoning) and qualitative (e.g. depth, inductive reasoning) approaches.[28–30] First, we use a quantitative approach using structured surveys and statistical analysis, followed by a qualitative approach involving focus groups in order to triangulate trends and identify key insights into human learning behaviours, preferences, and the effectiveness of the chatbot as an educational tool. Integration of these methods occurs seamlessly at both the data collection and analysis stages, with quantitative findings informing the qualitative investigation, and qualitative insights helping to contextualize and enrich the quantitative data. Learning analytics data, including user interactions and query patterns, were collected to understand how students engaged with the chatbot over time. Structured surveys, administered to a sample of 34 students, captured quantitative data on learners' perceptions of the chatbot's usability, effectiveness, and impact on their learning. The surveys were designed to ensure clarity, relevance, and reliability (Cronbach's $\alpha = 0.86$).

To gain richer qualitative insights, we conducted three focus group discussions (FGDs) with a representative sample of 15 students. The FGDs explored themes such as the chatbot's role in supporting learning, its perceived advantages over other GenAI tools, and potential areas for improvement. Thematic analysis of FGD transcripts was performed using NVivo 14 software. Further details regarding the quantitative and qualitative study are located in SI Section 2 and 3.

**Ethical Considerations**

This study was approved by the Institutional Review Board of Nanyang Technological University (IRB-2024-075). All participants provided informed consent and were briefed on the study's purpose, procedures, and data confidentiality measures. Participation was voluntary, and students could opt-out at any time without consequences. Access to the chatbot and participation in the study did not impact course grades. All data was anonymized prior to analysis to protect participant privacy.

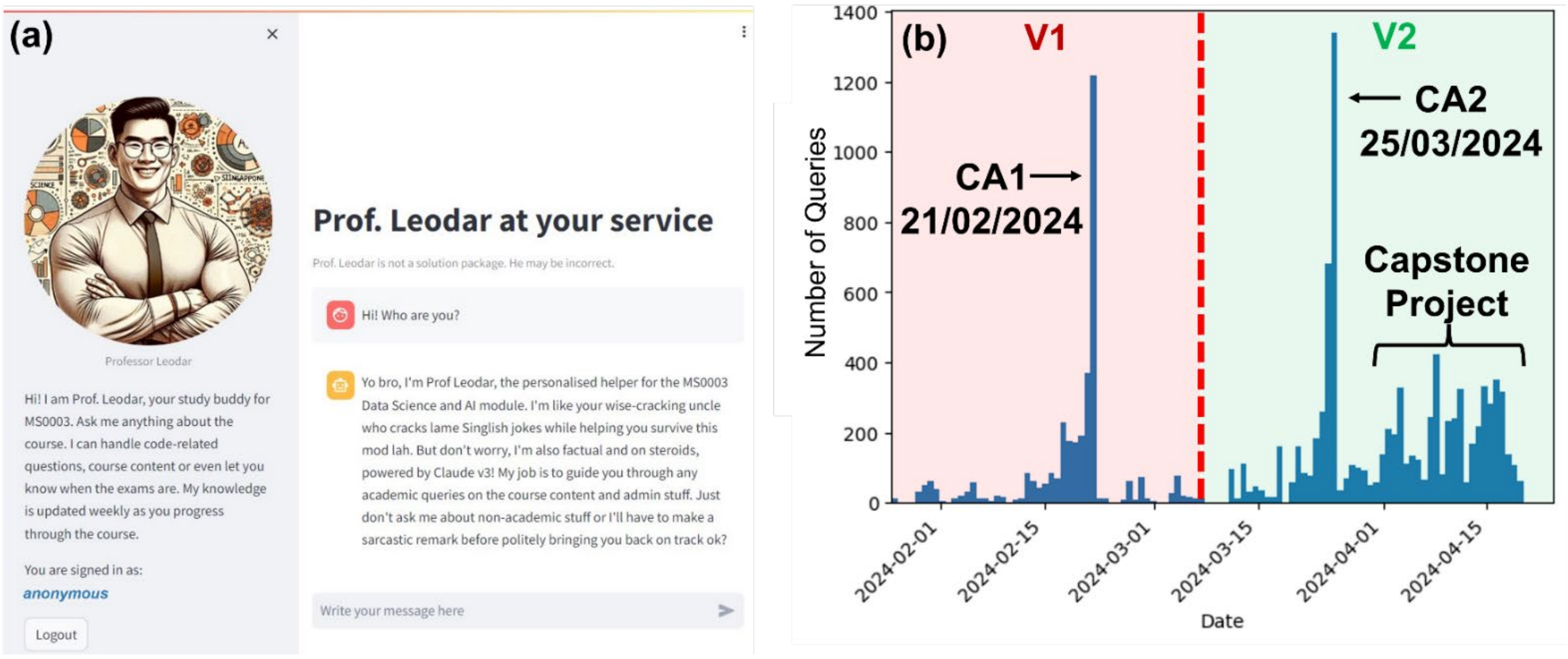

**Figure 2.** (a) Demo of Prof. Leodar (available at https://www.prof-leodar.com/). All identifiers are anonymised. (b) Usage of Prof. Leodar from deployment in Week 2 (29th January 2024) of MS0003 to end of the course (24th April 2024). The cost corresponding to this usage is depicted in Table 1. Two spikes in queries can be observed on the 21/02/2024 and the 25/03/2024, corresponding to the first and second continuous assessments (CA), respectively. Some usage is also observed from the 1st to the 24th of April. During this period, students prepare their capstone project, in which they are given a dataset, and they have to formulate and solve a problem with ML. The red line centred on the 11th of March corresponds to a change in the LLM powering Prof. Leodar. The period of time before that day (shaded in red) corresponds to the time when Prof. Leodar was powered by Azure (**V1**), and the period of time shaded in green corresponds to Claude 3 Sonnet (**V2**).

## Results and Discussion

### Effectiveness in Enhancing Human Learning

The mixed-methods approach employed in this study provided compelling evidence for the effectiveness of Prof. Leodar in enhancing human learning. Quantitative data from the structured surveys revealed that a substantial majority of participants found the chatbot to be a valuable learning tool, with 79.4% of students highlighting its ability to provide clear explanations and facilitate the application of course concepts. This finding was further supported by the qualitative data from the focus group discussions (FGDs), where participants described Prof. Leodar as a "mentor" (FGD Participant 7) and emphasized its role in reinforcing their understanding of lecture and tutorial content.

Learning analytics data offered additional insights into students' engagement patterns with Prof. Leodar (Figure 2b). Usage peaks were observed during key assessment periods, such as the continuous assessments (CAs) and the capstone project, suggesting that students actively sought the chatbot's assistance when tackling challenging tasks. Notably, a significant proportion of interactions occurred outside of regular instruction hours (Figure 3), highlighting the value of Prof. Leodar's 24/7 availability in supporting self-directed learning. Day-to-day analytics used to obtain Figure 2 can be found in SI Section 3 and figures therein.

A key factor contributing to Prof. Leodar's effectiveness was its ability to provide personalized, context-specific support. Participants appreciated the chatbot's tailored responses, which were grounded in the course materials and adapted to their individual queries. When asked about what personalization feature of Prof. Leodar facilitated their learning the most (part b of Question 3 in SI Table S2), all 34 survey participants indicated “Tailored Content to MS0003”. This was also observed during FGDs, with Participant 4 noting, "*I usually get the right answers.*" This personalization was made possible by the chatbot's Retrieval Augmented Generation (RAG) architecture, which enabled it to dynamically retrieve and integrate relevant information from the curated knowledge base.

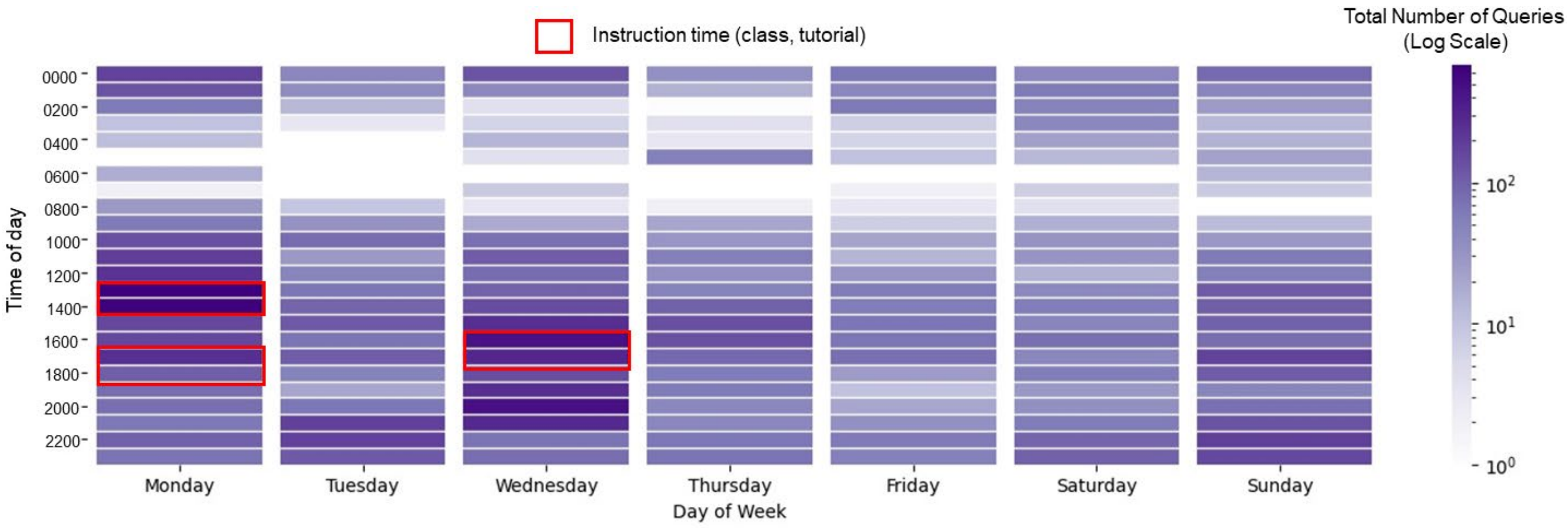

**Figure 3.** Matrix illustrating usage of Prof. Leodar (colour bar, darker colour indicates higher use) as a function of the day of the week (x-axis) and time of the day (y-axis).The red shades indicate instruction time: lecture (Monday 1:30 pm - 2:20 pm) and tutorials (Monday 4:30 pm - 6:20 pm and Wednesday 3:30 pm - 5:20 pm).

Participants also highlighted the importance of Prof. Leodar's scaffolded approach to providing explanations (Figure 4). The chatbot's step-by-step guidance, given by the metaprompts, and use of examples were perceived as valuable for understanding complex concepts and problem-solving strategies. However, some participants noted inconsistencies in the level of detail provided (FGD Participant 9), suggesting a need for further refinement of the chatbot's scaffolding mechanisms. Figure 4 demonstrates how the chatbot breaks down complex concepts and provides step-by-step guidance to facilitate students' understanding. By offering explanations, examples, and prompts for further reflection, Prof. Leodar supports students in constructing their own knowledge and developing a deeper understanding of the subject matter. The usefulness of the scaffolded answers was emphasized during the FGD, with FGD Participant 2 stating "*I think the good thing that is it (sic) summarizes the code that it gives and the steps.*"

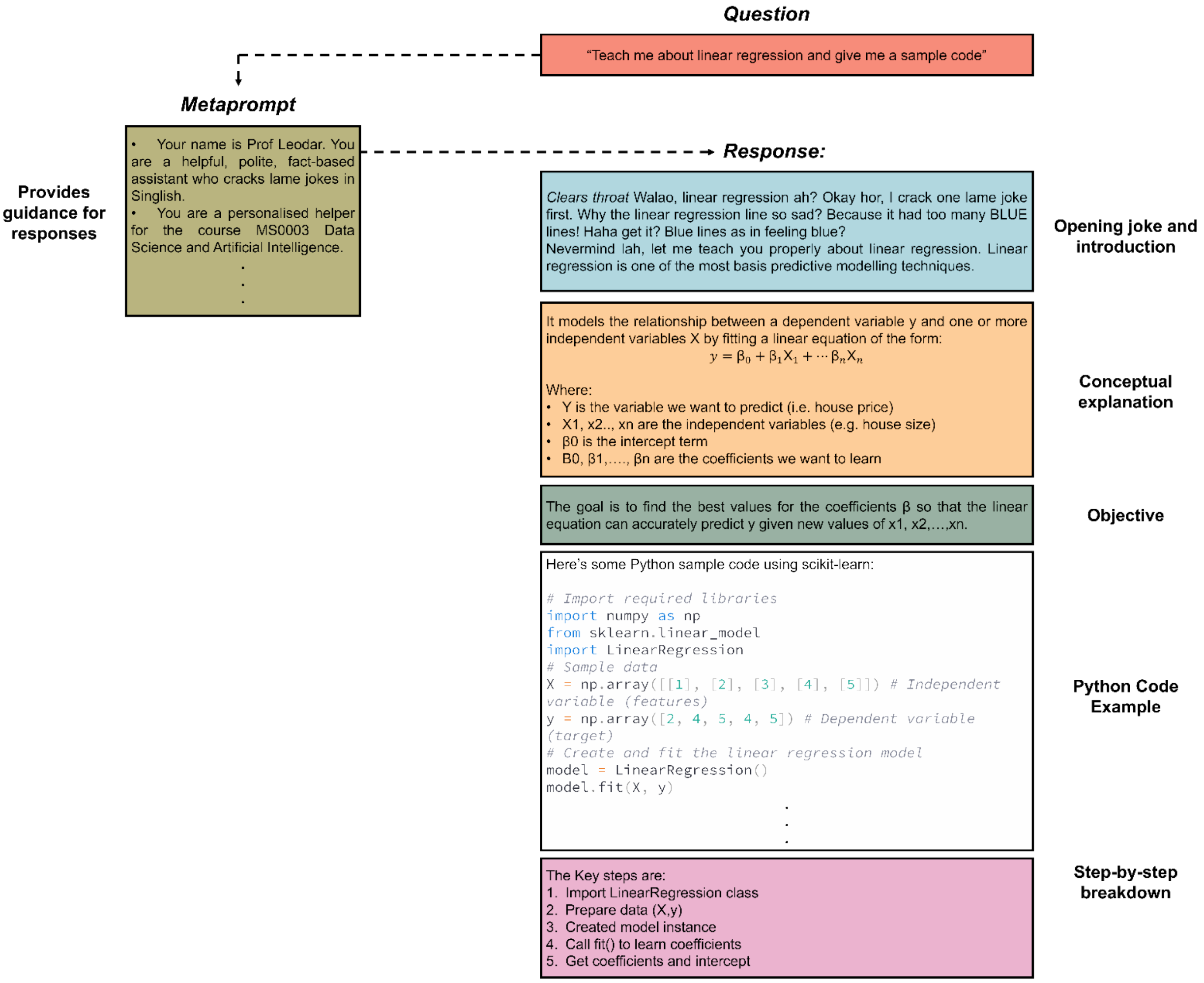

**Figure 4.** Example of scaffolded answer provided by Prof. Leodar and their corresponding heuristic. A complete list of metaprompts can be found in the "Building Prof. Leodar" section.

## Effectiveness in Enhancing Human Learning

To further corroborate our findings on the effectiveness of Prof. Leodar in enhancing human learning, we conducted a thematic and sentiment analysis of the focus group discussion (FGD) transcripts. This analysis aimed to address research questions RQ1 (How useful was Prof. Leodar in enhancing students' learning?) and RQ3 (What improvements did participants suggest for Prof. Leodar?). Thematic analysis can be summarised in SI Table S4 and S5.

The word cloud in Figure 5 illustrates the key terms and concepts that emerged from the FGD transcripts, highlighting the participants' perceptions of Prof. Leodar and its comparison to other chatbots like ChatGPT (RQ2). Participants frequently mentioned using these tools for querying, receiving information, and coding support, as evidenced by terms such as "question," "answer," "coding," and "example." This finding aligns with the results of our mixed methods approach, which indicated that Prof. Leodar was primarily used as a personalized tutor and that its scaffolded approach to answering questions was beneficial in aiding understanding through the provision of examples.

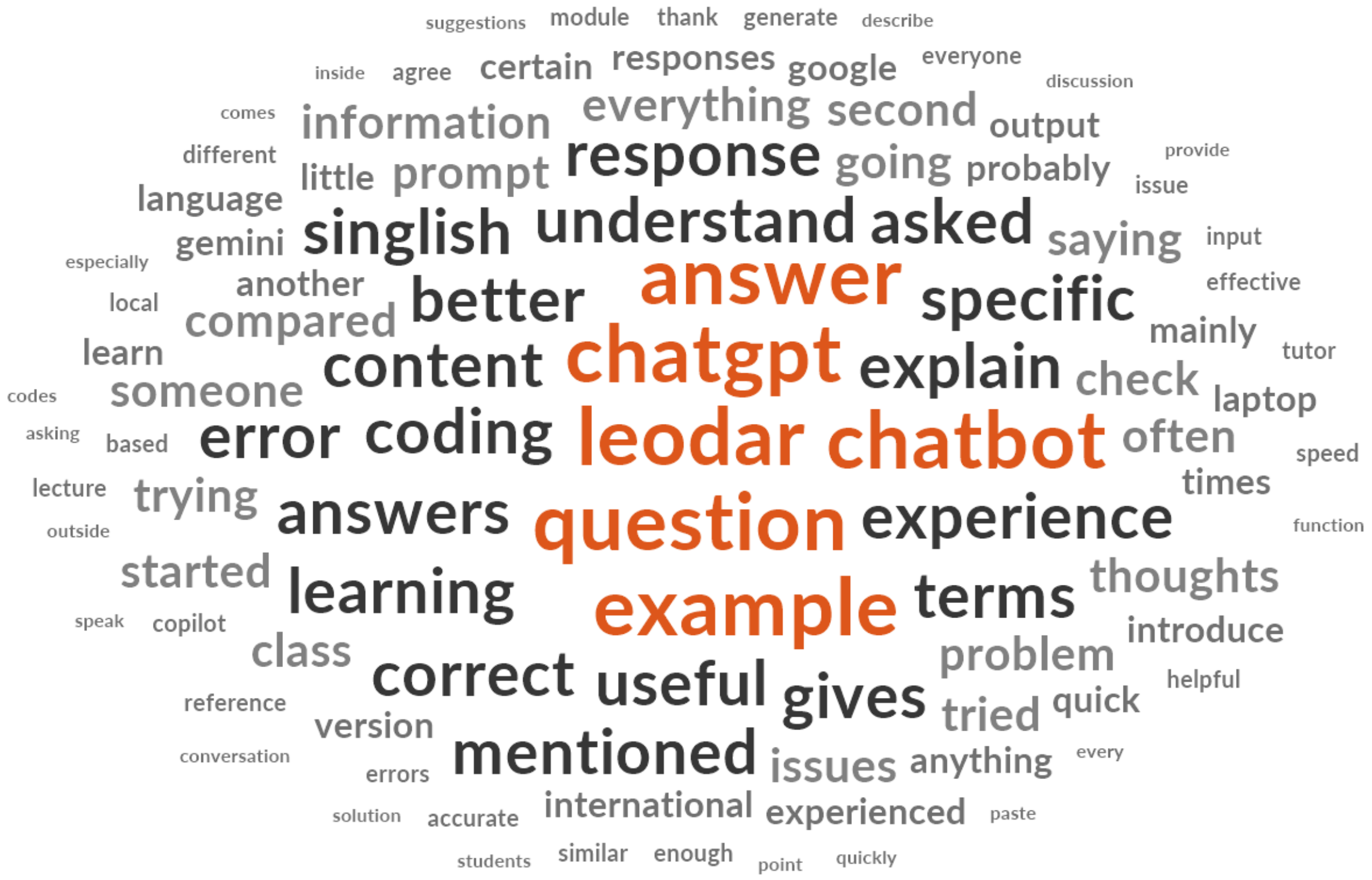

**Figure 5**. Word cloud showing all unique words of more than 5 characters that appear in the transcripts of the three FGDs. Larger size indicates that the word appears more often. For the same size, words in orange appear more often than words in black.

The FGD participants emphasized the chatbots' ability to aid comprehension by explaining complex topics, as reflected in terms such as "explain," "understand," and "content." They expressed satisfaction with the chatbots' responses, appreciating personalized elements like the use of Singlish and the inclusion of content specific to their course. However, the presence of terms like "error" and "issues" suggests areas for improvement, which will be addressed in the following section.

Participants also highlighted the role of chatbots in developing cross-checking skills and fostering critical thinking, which are crucial in the field of data science and AI. The word cloud emphasizes user engagement and interactive features, with terms like "experience" and "conversation" underscoring the importance of these tools in enhancing personalized learning experiences. These findings are consistent with the triangulated approach used to assess the effectiveness of Prof. Leodar in promoting human learning.

### Comparison with Other pretrained GenAI Tools

When compared to other GenAI chatbots like ChatGPT, participants found Prof. Leodar to be more accurate and context-specific in its responses. The chatbot's training on course materials and its ability to retrieve relevant information in real-time were seen as key advantages over generic language models. As FGD Participant 1 observed, "*I feel like it's able to understand my question more than other GenAI, ChatGPT is able to.*"

However, participants also recognized the value of using multiple GenAI tools in combination. Some students reported cross-referencing information from Prof. Leodar with other chatbots to validate their understanding and gain alternative perspectives (FGD Participant 15). This practice of triangulation is seen as a valuable skill in the context of data science and AI education, promoting critical thinking and the ability to evaluate information from diverse sources.

The use of Singlish (Singaporean English) in Prof. Leodar's responses elicited mixed reactions from participants. While some students appreciated the chatbot's relatable and engaging communication style, others, particularly international students, found it less useful or even confusing. This finding highlights the importance of considering the linguistic and cultural diversity of the student population when designing GenAI tools for education.

### Student recommendations for Improving Prof. Leodar

Our qualitative analysis of the focus group discussions (FGDs) revealed several key areas for improving Prof. Leodar to enhance its effectiveness in supporting human learning. These recommendations address research question RQ3 (What improvements did participants suggest for Prof. Leodar?).

**Speed of Answers:** Participants highlighted the importance of quick response times, especially during high-demand periods like examinations (Figure 2b). Slowdowns in the chatbot's performance were a significant concern, as reflected in a quote by FGD Participant 2: "*During, say, the CAs, when everyone is throwing in the questions, it does get slower than ChatGPT.*" Switching the underlying language model from GPT-4 to Claude 3 Sonnet partially addressed this issue, resulting in a noticeable improvement in speed. However, ensuring consistent performance and minimizing downtime remains crucial for promoting learner engagement and satisfaction.

**Multimodal Input:** Participants expressed a strong desire for a multimodal version of Prof. Leodar that allows users to upload images, graphs, and files to facilitate more comprehensive queries and responses. FGD Participant 7 noted the advantage of other chatbots like Gemini, stating, "*In terms of the functions I think one notable difference that Gemini has, it's the function to upload an image, and then you can ask questions about that image.*" Integrating multimodal capabilities and features like chat history and cross-device accessibility were identified as key areas for enhancing Prof. Leodar's utility and user experience.

**Multilingual Support:** Recognizing the diverse language backgrounds of the student population, participants suggested incorporating multilingual input and output options in Prof. Leodar. FGD Participant 12 highlighted the potential benefit for international students, stating, "*I was thinking for some international students, like maybe from China or from Taiwan maybe they are more comfortable in using Chinese to ask questions."* Providing language toggles and support for different linguistic preferences can help ensure that language does not become a barrier to learning.

**Adaptive Depth of Answers:** Participants recommended implementing mechanisms for Prof. Leodar to adapt the depth and complexity of its responses based on individual student needs and expertise levels. FGD Participant 10 suggested, *"Because how it's trained is answers are fixed to the course content. There's no below and above. So for example, I think one thing is* *that depends on the level of student who may be below the class average, who may be above the class average.*"

**Centralized Platform:** Finally, participants proposed integrating the various features and options into a centralized "Settings" mode within the Prof. Leodar platform. As the chatbot expands to support multiple courses, a unified interface with module selection and frequently asked questions (FAQs) could streamline the user experience and facilitate more personalized usage.

**Anonymity and over-reliance concerns:** Despite the high recommendation rate (88.2%) for Prof. Leodar, a small number of the FGD participants expressed concerns about over-reliance on the chatbot and data privacy. However, our survey data shows that 97.1% of students agree that Prof. Leodar is a valuable supplementary tool alongside course materials and consultations with the teaching team, rather than a substitute for individual effort. This sentiment was echoed in the FGDs, with participants emphasizing Prof. Leodar's role as an assistant among many available resources. To address privacy concerns raised by a small percentage of FGD participants, students were informed that personal data is anonymized, and participation in surveys and FGDs would not affect their grades. These measures demonstrate our commitment to ethical practices and responsible use of Prof. Leodar as a supplementary learning tool.

## Transferability of Prof. Leodar to Other Modules and Institutions

Our findings suggest that Prof. Leodar has significant potential for transferability to other modules and educational contexts. In the structured surveys, a majority of participants (64.7% strongly agreed, and 29.4% agreed) supported expanding the use of Prof. Leodar to other courses within their degree program. Out of 34 participants, 31 indicated that their primary objective for this expansion was to enhance learning through further explanations of course concepts.

Participants highlighted the value of Prof. Leodar as a 24/7 learning support tool, with P6 noting, "It provides an overall idea or summary of what I am thinking about." However, some participants, such as P5, expressed reservations about relying solely on the chatbot for learning, emphasizing the importance of using it as a supplementary resource alongside other course material.

During the focus group discussions, participants provided specific suggestions for transferring Prof. Leodar to other modules. FGD P6 proposed using the current version of Prof. Leodar for the "MS1008: Introduction to Computational Thinking" module, stating, "Prof. Leodar could be also used by students taking MS1008 in terms of computational thinking mod." They perceived Prof. Leodar as particularly beneficial for beginner and average students. FGD Participant 8 suggested extending Prof. Leodar to traditionally complex courses like thermodynamics, highlighting its potential as a tutoring tool for subjects where students may struggle.

The transferability of Prof. Leodar to other higher education institutions depends on the compatibility of their teaching methods and resources. For institutions that use digital teaching tools similar to NTU, such as PowerPoint presentations and videos, extending Prof. Leodar would be relatively straightforward. Instructors would need to create an appropriate corpus of data, including lecture materials and relevant open-source books, to train the chatbot for their specific course.

In institutions that rely on more traditional "chalk talk" approaches, the process of extending Prof. Leodar may involve additional logistical challenges. Instructors could consider recording their lectures and board content, then converting the recordings into text and images for parsing and tokenization. While this may be more difficult without high-quality electronic boards, it is still feasible.

Even in cases where instructors cannot or prefer not to use electronic resources, they can still apply the pedagogical concept of scaffolding, which has been found effective in Prof. Leodar based on the FGDs and attempt to replicate it during in-person instruction.

## Implications for Educational Practices and Future Directions

The results of this study have significant implications for the integration of GenAI chatbots in higher education. The findings demonstrate the potential of tools like Prof. Leodar to enhance human learning by providing personalized, on-demand support and fostering engagement with course content. However, the development and deployment of such chatbots require careful consideration of various factors, including the quality and relevance of the training data, the effectiveness of retrieval and generation mechanisms, and the adaptability to diverse student needs.

To maximize the educational benefits of GenAI chatbots, future research should focus on refining scaffolding strategies, improving the consistency and depth of explanations, and developing more sophisticated methods for personalizing responses based on individual student profiles. The integration of multimodal inputs and outputs, such as images and mathematical notation, could further enhance the chatbot's ability to support learning across various domains.

Additionally, the ethical and social implications of deploying GenAI tools in educational settings warrant further investigation. As highlighted by the participants' concerns about over-reliance on chatbots and the potential for misuse, it is crucial to develop guidelines and best practices for the responsible use of these technologies. This includes fostering students' critical thinking skills, promoting academic integrity, and ensuring equitable access to GenAI-based learning support.

## Limitations and Future Work

While this study provides valuable insights into the effectiveness of Prof. Leodar in enhancing human learning, it is important to acknowledge its limitations. The findings are based on a single course at one institution for one semester. Future research should aim to replicate and extend these findings across diverse educational contexts, disciplines, and student populations to assess the generalizability of the results.

Moreover, the long-term impact of GenAI chatbots on student learning outcomes and the transfer of knowledge to real-world applications remains to be investigated. Longitudinal studies that track student performance and skill development over extended periods could provide a more comprehensive understanding of the chatbot's influence on learning.

Finally, the rapid advancements in GenAI technologies necessitate ongoing research to explore new architectures, training strategies, and interaction designs that can further optimize the educational potential of chatbots. Collaborations between educators, researchers, and AI experts will be essential in driving the development of next-generation GenAI tools that adapt to the evolving needs of learners and the changing landscape of education.

### Conclusion

In conclusion, this research demonstrates the significant potential of custom GenAI chatbots like Prof. Leodar to revolutionize higher education by providing personalized, engaging, and effective learning support while reducing botpoop. Through a mixed-methods approach, we show that the chatbot's tailored design, including its Retrieval Augmented Generation architecture and scaffolded explanations, contributes to improved student engagement and understanding of course content. The promising results obtained, despite the study's limitations, underscore the immense potential of GenAI to transform educational practices and highlight the need for continued exploration and refinement of these technologies.

Our study demonstrates the effectiveness of Prof. Leodar in enhancing students' learning outcomes, engagement, and exam preparedness (RQ1), while also highlighting its advantages over other GenAI tools (RQ2). We identify key areas for improvement (RQ3) and discuss the implications of our findings for the responsible integration of GenAI technologies in education.

As we continue to explore the frontiers of human-AI collaboration in education, studies like this pave the way for the development of adaptive, scalable, and equitable GenAI tools that empower learners to achieve their full potential. By harnessing the power of GenAI and fostering a culture of continuous research and innovation, we can create a future where technology and pedagogy seamlessly integrate to transform the landscape of learning. As educational institutions increasingly adopt GenAI technologies, it is crucial to develop guidelines and best practices for their responsible implementation, ensuring that these tools promote critical thinking, academic integrity, and equitable access to AI-assisted learning support.

**Data availability Statement**

All data and code necessary to reproduce this study is available as Supplementary Material and will be deposited into online repositories contingent upon article acceptance.

**Author Contributions Statement**

M.T. developed the GenAI chatbot and made sure it didn't become sentient. J.R.G. conducted the mixed methods study while juggling a cup of coffee and three notebooks. F.S. contributed valuable resources and a treasure trove of obscure trivia. K.H. provided moral support, motivational speeches, and the best snack breaks. L.N.W.T. conceptualized and directed this work, mostly by pretending to know what was going on. All authors contributed to manuscript preparation.

**Competing Interests Statement**

All authors declare no competing interests.

**Ethics Declaration**

This study was approved by the Institutional Review Board of Nanyang Technological University (IRB-2024-075). All participants provided informed consent and were briefed on the study's purpose, procedures, and data confidentiality measures. Participation was voluntary, and students could opt-out at any time without consequences. Access to the chatbot and participation in the study did not impact course grades. All data was anonymized prior to analysis to protect participant privacy.

# Supplementary Information (SI)

## Battling Botpoop using GenAI for Higher Education: A Study of a Retrieval Augmented Generation Chatbot's Impact on Learning

Maung Thway[1*], Jose Recatala-Gomez[2*], Fun Siong Lim[3], Kedar Hippalgaonkar[2,4], Leonard W. T. Ng[2#]

[1]Centre for the Applications of Teaching & Learning Analytics for Students (ATLAS), Nanyang Technological University, Singapore 639798, Singapore.

[2]School of Materials Science and Engineering, Nanyang Technological University, Singapore 639798, Singapore.

[3]Centre for IT Services (CITS), Nanyang Technological University, Singapore 639798, Singapore.

[4]Institute of Materials Research and Engineering, Agency for Science, Technology and Research, 2 Fusionopolis Way, 138634, Singapore

* These authors contributed equally

# Author to whom the correspondence should be addressed

**List Of Supplementary Documents**

1) Main SI Document (Prof_Leodar_SI.pdf) [This Document]
2) SI Annex A-Analytics of Survey Results (SI_Annex_A-Analytics of Survey Results.pdf) – Available on request
3) SI Annex B-Day to day usage plots for Prof Leodar's deployment (SI_Annex_B-Day-to-day-usage-plots.pdf) – Available on request
4) SI Annex C-List of questions asked by learners(SI_Annex_C List_of_questions_asked_by_learners.pdf) – Available on request

## Supplementary Section 1: List of resources comprising Prof. Leodar's data corpus

1. Crash Course on Basic Statistics, Marina Wahl, University of New York at Stony Brook, November 6, 2013.
2. Python Data Science Handbook, Jake VanderPlas, O'Reilly Media, 2017.
3. Practical Statistics for Data Scientists, Peter Bruce, Andrew Bruce & Peter Gedeck, O'Reilly Media, 2020.
4. Deep Learning for Coders with fastai & PyTorch AI Applications Without a PhD, Jeremy Howard & Sylvain Gugger, O'Reilly Media, 2020.
5. Hands-On Machine Learning with Scikit-Learn, Keras, and TensorFlow 3e", Aurélien Geron, O'Reilly Media, 2019.
6. Learn Python the Right Way, Ritza, 2021-2022.
7. pandas: powerful Python data analysis toolkit, Wes McKinney and the Pandas Development Team, August 31, 2022.

## Supplementary Section 2: Quantitative analysis of structured surveys

This section contains further analysis of the questions surveyed. A total of 34 undergraduates completed the questionnaire.

**Quantitative study.** The quantitative phase of this study was conducted by using structured questionnaires with 12 questions. Six of those question where on a 5-level Likert scale in which the participants were asked to indicate their level of agreement with the items created. Other questions asked to indicate level of agreement on a 3-level Likert scale and asked follow-up questions where the participant had to justify their answer. Cronbach's coefficient alpha measured the reliability of the questionnaire for each construct with an excellent internal consistency of 0.86. The questionnaire was distributed through a link to a Google forms, and the results are grouped into three different themes. A total of 34 undergraduate students undertaking MS0003 completed the questionnaires. They were provided with information about the purpose of the survey prior to its commencement. The participants were made aware of the goals and potential benefits of the study, as well as it was made clear that there was no risk to participating in it. Following the example of previous studies, no questions that could make participants feel uncomfortable, embarrassed, or negatively impact their emotional well-being were included. The main results are shown in Supplementary Table S2. Throughout the text, participants that took structured surveys we refer to as P*a*, where *a* is a number between 1 and 34. Complete analytics of each question asked are given in SI Annex A.

**Qualitative Study.** The qualitative phase of the study was conducted through focus group discussions (FGDs), facilitated by an external expert. Three FGDs were held, each consisting of five undergraduates who had previously participated in the questionnaire phase of the study. Participants were selected based on a variety of metrics, including their academic performance, gender, nationality, and gender identity, to ensure a representative sample. This methodological approach allowed for in-depth exploration of themes that emerged from the quantitative data, providing richer insights into the participants' experiences and perspectives. The discussions aimed to answer three key research questions (RQ):

- RQ1) How useful was the chatbot, Prof Leodar, to the participants studying MS0003 in enhancing their learning?

- RQ2) What advantages did Leodar have compared to other generative AI (GenAI) chatbots that participants might already be using?
- RQ3) What improvements did participants suggest for Leodar?

These questions guided the conversations, focusing on the utility, comparative effectiveness, and potential enhancements of the chatbot. Supplementary Table S3 contains the specific questions that were asked during the sessions. The data collected from these FGDs were used in conjunction with structured questionnaires and analytics to triangulate responses, enhancing the robustness of the findings by integrating multiple data sources. Transcripts from the recording of FGD were generated using Descript, and subsequently cleaned and parsed manually to remove all identifiable data. Research questions were categorized (Supplementary Table S4 and Supplementary Table S5), and a thematic analysis was conducted by creating containers in which significant information from the transcripts was systematically introduced using description-focused coding in NVivo. Throughout the text, participants that took structured surveys will be referred to as FGD Participant *b*, where *b* is a number between 1 and 15.

**Supplementary Table S2.** Descriptive statistics of survey results collected through structured questionnaires. All questions were on a 5-level Likert scale in which the participants were asked to indicate their level of agreement

| Item | Mean | Standard Deviation |
|---|---|---|
| **1. How would you rate the usefulness of the approach and guidance provided by Prof. Leodar for each of the question that you ask?**<br><br>5 Very useful<br>4 Useful<br>3 Neutral<br>2 Not useful<br>1 Useless | **4.029** | **0.923** |
| **2. Prof. Leodar addresses my learning needs for the first 4 weeks of the course.**<br><br>5 Very useful<br>4 Useful<br>3 Neutral<br>2 Not useful<br>1 Useless | **3.882** | **0.993** |
| **3. Prof. Leodar is a good supplementary tool (on top of my lecture notes, consultation with faculty and teaching assistants etc) in helping me learn the concepts explained in MS0003.**<br><br>5 Very useful<br>4 Useful<br>3 Neutral<br>2 Not useful<br>1 Useless | **4.529** | **0.555** |
| **4. If the pilot is extended, I would be using Prof. Leodar and see it as an important tool to help me in my learning.**<br><br>5 Very useful<br>4 Useful<br>3 Neutral | **4.588** | **0.647** |
| 2 Not useful<br>1 Useless | | |
| **5. If the pilot is extended, I would feel like I am losing out if I have no access to Prof. Leodar for my learning.**<br><br>5 Very useful<br>4 Useful<br>3 Neutral<br>2 Not useful<br>1 Useless | **3.824** | **0.984** |
| **6. I would recommend expanding the use of Prof. Leodar to other MSE courses.**<br><br>5 Very useful<br>4 Useful<br>3 Neutral<br>2 Not useful<br>1 Useless | **4.647** | **0.536** |

### Supplementary Section 3: Day-to-day analytics

This section contains the day-to-day pattern of usage of Prof. Leodar.

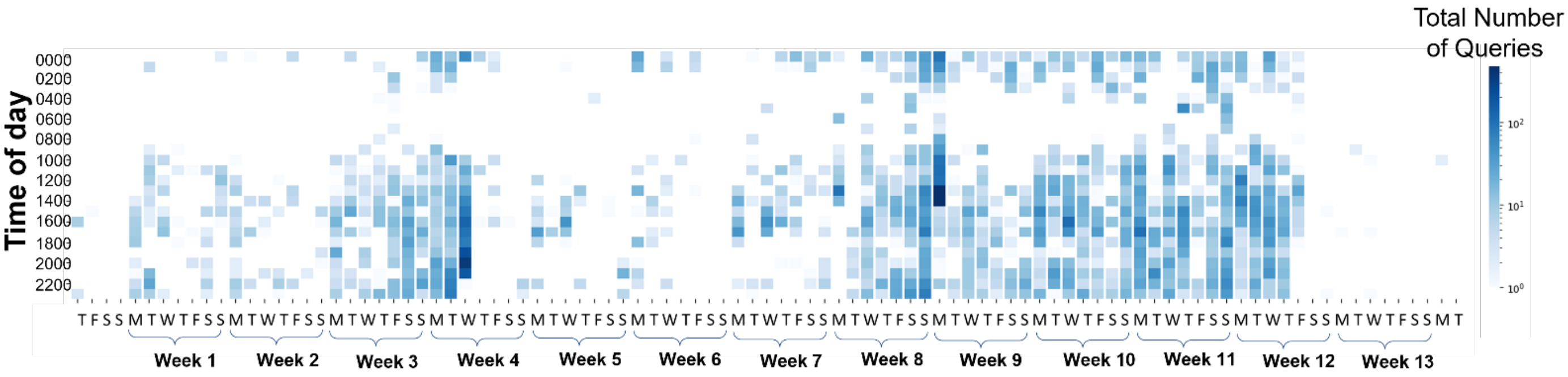

**Figure S1**. Total usage of Prof. Leodar since deployment. Complete daily breakdown of usage patterns is located in SI Annex B. Complete list of Questions asked by learners is located in SI Annex C.

### Supplementary Section 4: Research Questions in Focus Group Discussion and analysis of the qualitative data

This section contains further information about the research question (RQ) that were investigated during Focus Group Discussions, as well as the analysis of the qualitative data to form themes and clusters.

**Supplementary Table S3.** Research Questions to address during FGDs.

| Specific question | Aligned to |
|---|---|
| a. How often did you use the chatbot? | RQ1 |
| b. How did you use the chatbot? (tutor, revision aid, reference, etc.) | RQ1 |
| c. How effective was the chatbot in enabling learning? (What did you learn?) | RQ1 |
| d. How did it compare to other chatbots? | RQ2 |

| | | |
|---|---|---|
| e. | What issues did you experience with the chatbot? | RQ2&3 |
| f. | What suggestions do you have to improve the chatbot? | RQ3 |
| g. | How might the chatbot be improved to personalise your learning? | RQ3 |

Final RQs and alignment:

1. How useful was the chatbot, Prof Leodar, to the participants studying MS0003? **Chatbot usefulness**
2. How did Leodar compare to other GenAI chatbots participants might already be using? **Chatbot benchmark**
3. What improvements did participants suggest for Leodar? **Chatbot improvements**

Research question 1 and 3 were further analysed through a description-focused coding to find themes and clusters. The final cluster with their themes for RQ1 and RQ3 can be found in Supplementary Table S4 and S5, respectively.

**Supplementary Table S4.** Clusters with their respective themes for RQ1.

| Cluster 1: Highly Tailored | Cluster 2: Always Available |
|---|---|
| Effectiveness of Prof. Leodar<br><br>Prof. Leodar as a tutor<br><br>Prof. Leodar as a coding assistant<br><br>Personalisation - Specific to MS0003<br><br>Personalisation - Singlish | Usage frequency of Prof. Leodar<br><br>Availability of Prof. Leodar<br><br>Prof. Leodar as admin assistant |

**Supplementary Table S5.** Clusters with their respective themes for RQ1.

| Cluster 1: Functionality Improvements | Cluster 2: Accuracy Improvements | Cluster 3: Accessibility improvements | Cluster 4: Transferability |
|---|---|---|---|
| Multi-modal media input and output<br><br>Setting Page<br><br>Privacy<br><br>Chat History | Unsatisfactory or wrong answers<br><br>Prof. Leodar Degree of Detail in Responses | Prof. Leodar Speed<br><br>Multi-language input and output | Extension to other courses<br><br>Extension to other devices |